\documentclass[a4paper,11pt]{article}
\pdfoutput=1 

\usepackage{jcappub} 

\title{\boldmath Results from a Hidden Photon Dark Matter Search Using a Multi-Cathode Counter}

\author[1]{A. Kopylov,\note{Corresponding author.}}
\author{I. Orekhov,}
\author{V. Petukhov}

\affiliation{Institute for Nuclear Research RAS,\\prospect 60-letiya
Octyabrya 7a, Moscow 117312, Russia}

\emailAdd{beril@inr.ru}

\abstract{Here we present measurements of the rates of emission of
single electrons from the cathode of a proportional counter filled
with a mixture of Ar + CH4 (10\%) at 1 bar. We interpret the results
as a possible photoelectric effect associated with hidden photons
(HPs). Our results set upper limits for HPs from cold dark matter
(CDM). We also discuss future options for searches for HPs from CDM
using a multi-cathode counter technique.}

\begin{document}
\maketitle
\flushbottom

\section{Introduction}
\label{sec:intro}

To advance experimental efforts in the search for dark matter,
physicists are now looking for new approaches. One such approach is
to use a dish antenna to observe hidden photons (HPs) from cold dark
matter (CDM)~\cite{1}. This technique has placed an upper limit on
the HP mass of $m_{\gamma'} = 3.1 \pm 1.2$ eV~\cite{2}. This range
of masses is limited by the coefficient of reflection of light from
the mirror and by the spectral sensitivity of the photomultiplier
tube. At higher frequencies, the light is absorbed by the reflecting
surface, which results in a loss of sensitivity for this technique.
However, the absorption of higher-frequency light causes the
emission of electrons from the surface if the photon energy is
higher than the work function of the metallic reflector. Hence, by
counting single electrons emitted from the metal one can extend the
sensitivity of this technique into the vacuum ultraviolet range and
higher.

We have developed a special detector for this study---a
multi-cathode counter~\cite{3}---and we have previously reported the
first results of measurements using a counter with a copper
cathode~\cite{4}. We have used a new counter of improved design,
which employs an aluminium cathode and focusing rings, to make
measurements with increased sensitivity~\cite{5}. Here we summarise
the results of all of our measurements and outline our plans for
further research.

\section{Measurement strategy}
\label{sec:meas}

If dark matter is composed entirely of HPs, then the power collected
by a detector (here, by a cathode of the counter) is

\begin{equation}
\label{eq:1}
\begin{split}
{P=2\alpha^{2}\chi^{2}\rho_{CDM}A_{cath}}
\end{split}
\end{equation}

\noindent where $\alpha=cos(\theta)$ and $\theta $ is the angle
between the direction of the HP field, when it points in the same
direction everywhere, and the surface of the cathode, with $\alpha^2
= 2/3$ if the HP vector is distributed randomly; $\rho_{CDM} \approx
0.3 GeV/cm^3$ is the energy density of CDM, which we assume to be
equal to the energy density of the HPs; $A_{cath}$ is the area of
the cathode of the counter; and $\chi$ is a dimensionless parameter
that quantifies the kinetic mixing as it is explained in~\cite{1}.
If this power is converted into single electrons that are emitted
from the cathode of the counter, then

\begin{equation}
\label{eq:2}
\begin{split}{P=m_{\gamma'} \cdot R_{MCC} /\eta}
\end{split}
\end{equation}

\noindent where $m_{\gamma'}$ is the mass (energy) of an HP, $\eta$
is the quantum efficiency for a photon with energy $m_{\gamma'}$ to
yield a single electron from the surface of the metal, and $R_{MCC}$
is the rate at which single electrons are emitted from the cathode,
which is presumed here to be entirely due to HPs. Thus, by combining
\eqref{eq:1} and \eqref{eq:2} we obtain:

\begin{equation}
\label{eq:3}
\begin{split}
\chi_{{\rm sens}}=2.9\cdot10^{-12}\left(\frac{R_{MCC}}{\eta\cdot 1
Hz}\right)^{\frac{1}{2}}\left(\frac{m_{\gamma'}}{1
eV}\right)^{\frac{1}{2}}\left(\frac{0.3\,{\rm
GeV/cm^3}}{\rho_{CDM}}\right)^{\frac{1}{2}}\left(\frac{1
m^{2}}{A_{MCC}}\right)^{\frac{1}{2}}\left(\frac{\sqrt{2/3}}{\alpha}\right)
\end{split}
\end{equation}

Here we assume the quantum efficiency $\eta$ for the conversion of
the HP at the surface of the metal cathode to be equal to the
quantum efficiency for a real photon of the same energy. The
sensitivity depends critically upon the dark rate of the counter.
Single electrons can be emitted by defects on the surfaces of the
wires, by protrusions, and by heterogeneous spots on the surface of
the cathode. Given these conditions, the value of $\chi$ obtained
from \eqref{eq:3} is only an upper limit. To improve the limit, we
need to decrease the dark rate of the counter. One way to accomplish
this would be to apply a surface treatment to diminish the effects
of extraneous sources. Another possibility would be to lower the
temperature of the detector to diminish the contribution from
thermionic emission, which depends on the work function $\phi_W$ of
the metal and the temperature \textit{T} in Kelvins, as expressed by
the Richardson equation:

\begin{equation}
\label{eq:4}
\begin{split}
R_{therm} =a\cdot T^2 \cdot e^{-\frac{\phi_W }{kT}}
\end{split}
\end{equation}

\noindent where $R_{therm}$ is the thermionic dark rate, \textit{a}
is a constant, and \textit{k} is Boltzmann's constant. Using an
$Ar/CH_4$ mixture enables us to lower the temperature to
$-40^{\circ}C$, which can substantially reduce the thermionic dark
rate.

\section{Apparatus}
\label{sec:appa}

\begin{figure}[htb]
\begin{center}\vspace{-0.2cm}
\includegraphics[width=10cm]{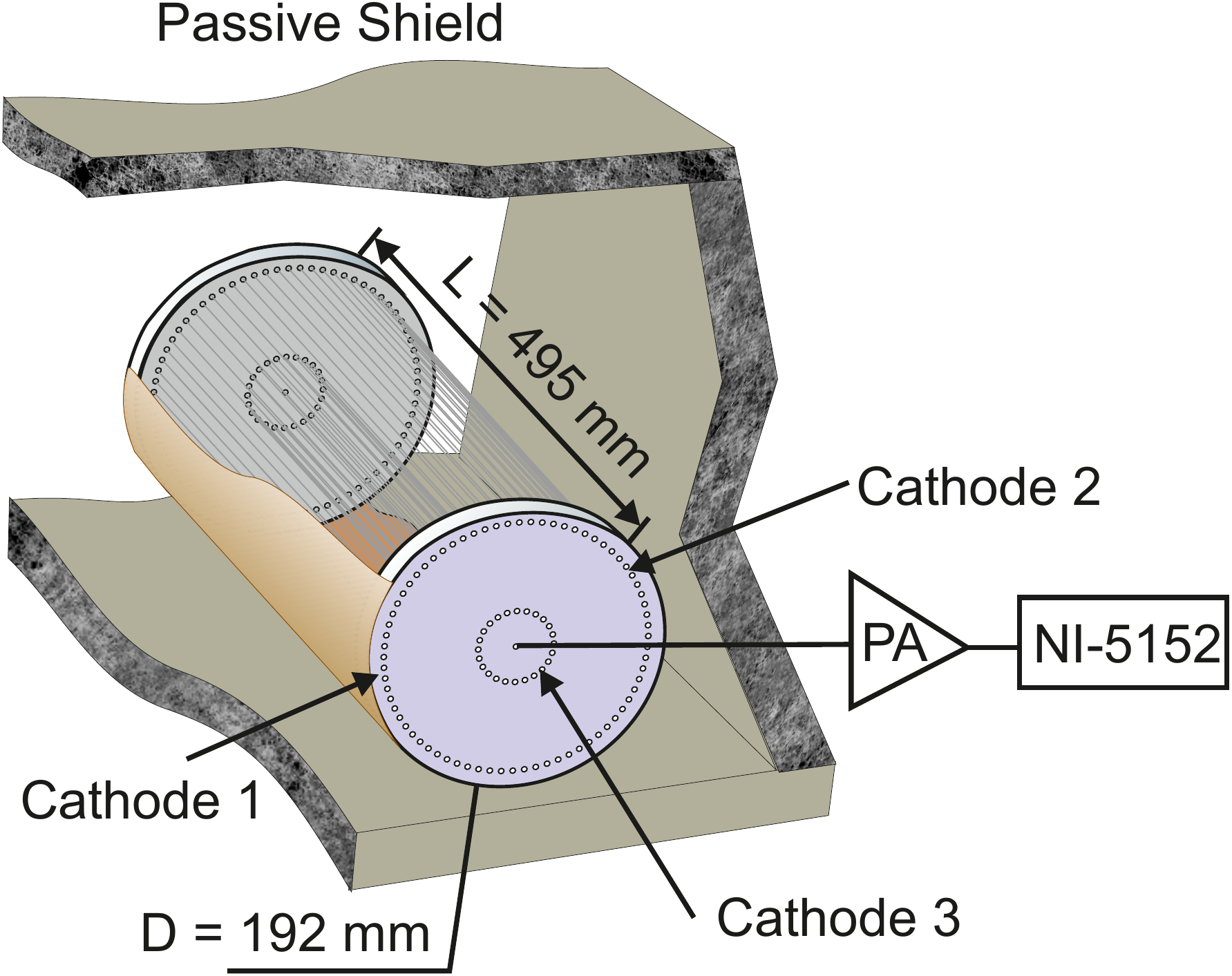}
\caption{\label{fig1} Schematic view of the multi-cathode counter in
the shielded cabinet.}
\end{center}
\end{figure}

Figure~\ref{fig1} shows a schematic illustration of a multi-cathode
counter. The counter is filled with an Ar + CH4 (10\%) mixture at 1
bar. It has three cathodes and one anode made of gold-plated
tungsten-rhenium wire 25 $\mu m$ in diameter. We apply the low
(negative) potential HV1 to cathode 1, the outer cathode, which is
fabricated from a metallic sheet, and we measure the counting rate
of individual electrons emitted from the surface of this sheet. The
second cathode is placed 5 mm from the outer cathode. It consists of
a series of nichrome wires 50 $\mu m$ in diameter, with a pitch of
4.5 mm. This cathode serves as a barrier to the electrons emitted
from outer cathode. We apply the potential HV2 to cathode 2. In
configuration 1, HV2 is higher than HV1, so that electrons emitted
from cathode 1 can drift to the central counter with the
still-smaller-diameter cathode 3, which is made of nichrome wires 50
$\mu m$ in diameter, with a pitch of 6 mm. The smaller (40 mm)
diameter of cathode 3 is used to obtain the high ($\approx 10^5$)
gas amplification needed to register the signal from single
electrons. In configuration 2, the potential HV2 is lower than HV1,
so that electrons emitted from the outer cathode are scattered back
from cathode 2 and do not reach the central counter. We use this
configuration to measure the dark rate due to electrons emitted from
the wires and to ionising particles that cross the counter along
short tracks at both ends. The two configurations have similar
geometries and similar electric fields, so we expect that the
difference $R_1-R_2$ between the count rates measured in these two
configurations measures the net effect due to electrons emitted from
the outer cathode.

To reduce the background due to gamma radiation from the surrounding
walls, we placed the counter in a special cabinet made of steel
slabs 30 cm thick. We have measured the $\gamma$-ray spectrum both
inside and outside the shielding with a NaI(Tl)detector and the
attenuation was obtained by comparing the two spectra. The
attenuation of the gamma radiation by the shielding was greater than
a factor of 100 for gamma rays at the $\approx 200$ keV peak energy
of the gamma-spectrum measured by a NaI(Tl) crystal.  Measurements
with the copper counter showed that with the counter placed outside
the shield the count rate for single electrons was about 20\% higher
than the one obtained inside the shield. From here, we found that
for the copper counter inside the shield the contribution to the
single-electron count rate due to gamma radiation is less than 0.2\%
of the single-electron count rate. The given estimation of the
remaining contribution from ambient gammas is a rough one like an
order estimation. No events have been observed coincident with muons
crossing the detector. The rate $R_{MCC}$ is obtained here as the
difference $R_1-R_2$ between the count rates in the two
configurations, with different potentials at the second
cathode~\cite{3}; see Figure~\ref{fig1}. We assume that in
configuration 1, the count rate is determined by the rate of
emission of single electrons from the cathode of the counter, the
wires and particles with short tracks crossing the counter at both
ends (the end effect). In configuration 2, the rate is determined by
the rate of emission from the wires and the end effect. The exact
contributions of the end effect and the wires to the total count
rate in configuration 2 is unknown for this technique. This
uncertainty is the main source of our systematic error.

We estimated the number of electrons lost during diffusion from the
outer cathode of the counter to the central counter based on the
attachment mechanism proposed by Bloch and Bradbury and developed by
Herzenberg (BBH model)~\cite{6,7} for electrons of small energies.
The number of free electrons in a gas that contains electronegative
impurities (in our case, oxygen impurities) decreases exponentially:

\begin{equation}
\label{eq:5}
\begin{split}
N(t) = N(0) \cdot e^{-At}
\end{split}
\end{equation}

\noindent where N(t) is the number of electrons at time t, N(0) is
the initial number of electrons, and A is the attachment rate. The
latter is given by the expression

\begin{equation}
\label{eq:6}
\begin{split}
A = P(M)\cdot P(O_2)\cdot C_{O_2,M}
\end{split}
\end{equation}

\noindent where $P(M)$, $P(O_2)$ are the pressure of the working gas
and of oxygen, respectively, and $C_{O_2,M}$ is the coefficient of
attachment, which does not depend upon the pressure of the gas or
the impurities in the BBH model.

\begin{figure}[htb]
\begin{center}\vspace{-0.2cm}
\includegraphics[width=10cm]{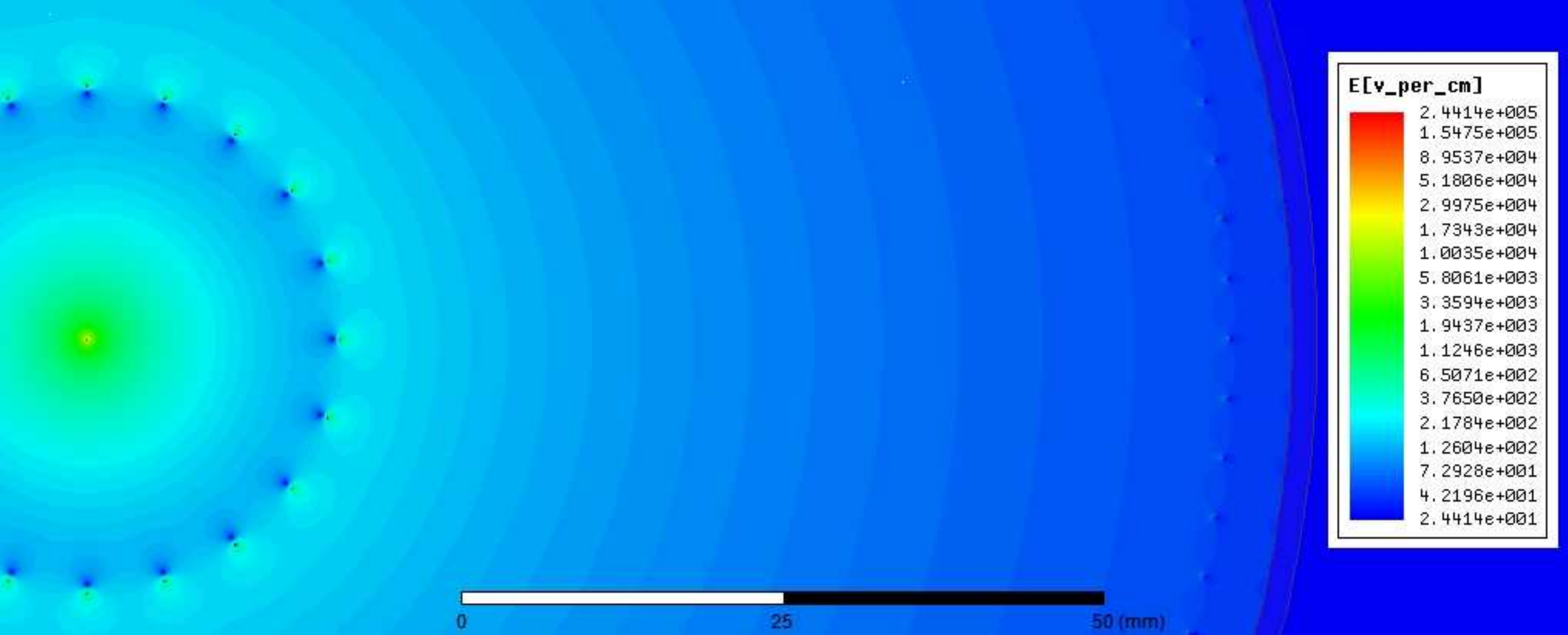}
\caption{\label{fig2} The calculated electric-field strength across
the counter.}
\end{center}
\end{figure}

The field strength in our detector, as calculated using Maxwell16,
is presented in figure~\ref{fig2}.

Using data for the dependence of the drift velocities of the
electrons upon E/p in the argon-methane mixture~\cite{8} and for the
coefficient of attachment $C_{O_2,M}$ from~\cite{9}, one can
determine the probability for an electron to become attached while
drifting from the outer cathode to the central counter. We found
that for our gas mixture this probability is less than 1\%;
\textit{i.e.} it can be neglected.

\section{Data analysis}
\label{sec:data}

We digitised the pulse shapes in intervals of $\pm$ 50 mV at a
frequency of 10 MHz and a sampling step of 400 $\mu$V. Each
measurement lasted for 12 h, after which we processed the data
offline. To select ``true'' events, we performed a selection in the
three-parameter space defined by the amplitude of the pulse, the
duration of the leading edge of the pulse, and a parameter $\beta $
that describes the prehistory of the event. The latter quantity is
proportional to the first derivative of the baseline, which we
approximated as a straight line during the 50 $\mu$sec before the
leading edge of the pulse. We estimated the efficiency as the
probability for the pulse to belong to the ROI box in this
3-parameter space; we found it to be (88 $\pm $ 6)\%. To reduce the
influence of noise on the counting, we selected only intervals with
a baseline deviation not more than 5 mV from zero, taking into
account a proper correction for the live time of the counting, which
we found to be about 54\%.

\begin{figure}[htb]
\begin{center}\vspace{-0.2cm}
\includegraphics[width=8cm]{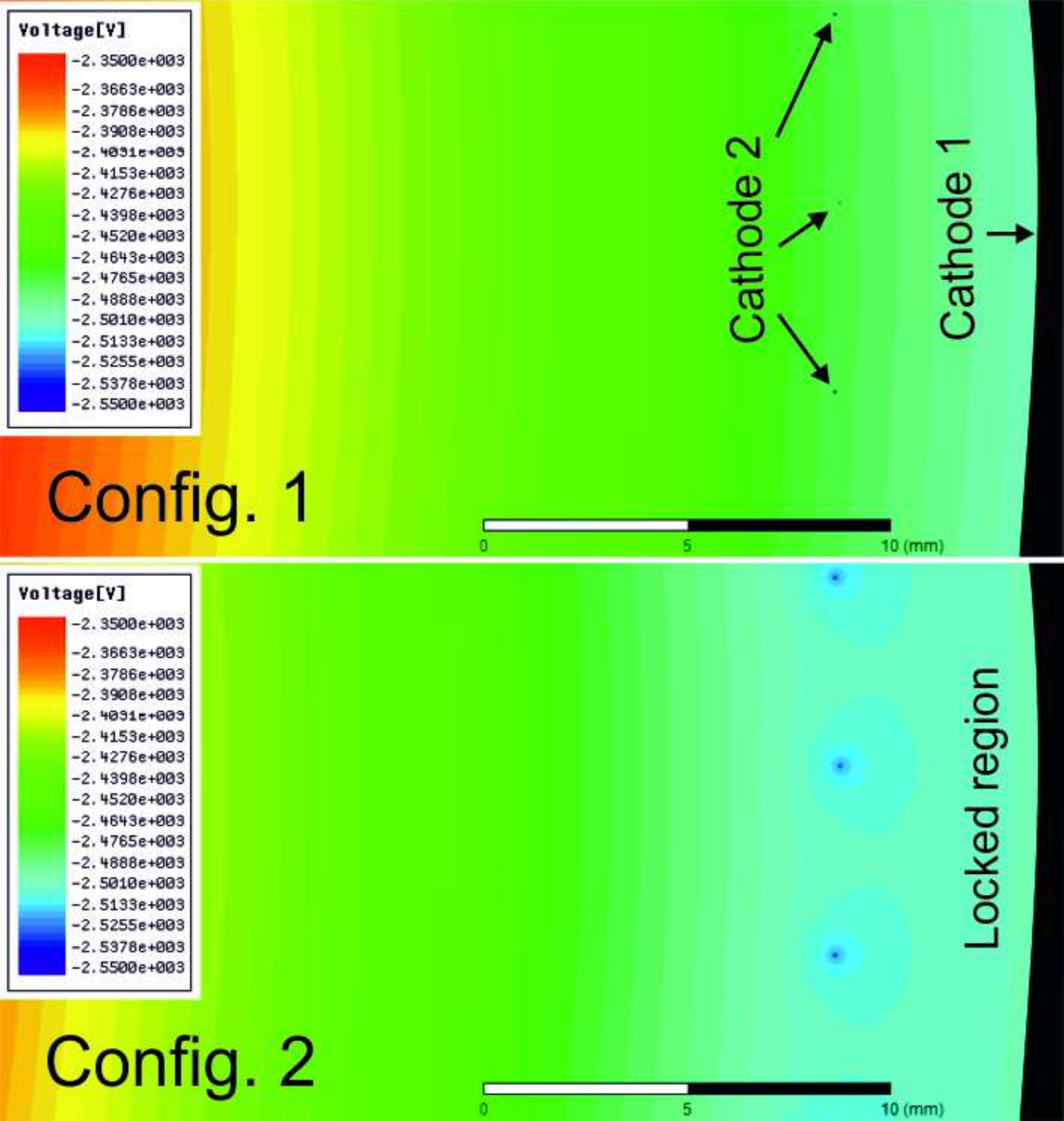}
\caption{\label{fig3} Calculated electric potentials across the
counter for the two configurations.}
\end{center}
\end{figure}

To evaluate the contribution to the total measured rate of the rate
of emission of electrons from the wires, we need to consider that in
configuration 2 only those parts of the surfaces of the wires that
face the centre of the counter produce the effect. Electrons emitted
from the opposite sides are retarded by the potential of the second
cathode and cannot drift toward the central counter. They are thus
trapped in the region between the first and second cathodes
(figure~\ref{fig3}). This reduces the background measured in
configuration 2. Instead of $R_1-R_2$, which is the appropriate
value if the entire rate is due to the end effect, in the case where
the entire rate is due to the wires we should use:

\begin{equation}
\label{eq:7}
\begin{split}
R_{MCC} =R_1 -\frac{n_3 +n_2}{n_3 +n_2/2} R_2
\end{split}
\end{equation}

Here $n_2$ and $n_3$ are the numbers of wires in the second and
third cathodes, respectively, see~\cite{5}.

\section{Summary}
\label{sec:sum}

In this section, we summarise the results of all our measurements,
using $R_1-R_2$ to evaluate the net event rate. We took the limit
Cu-1 from the results in~\cite{4}, which we obtained by using a
counter with a copper outer cathode at room temperature. At the
beginning of these measurements, we observed some instability in the
counter, which resulted in large scattering of the experimental data
points and consequently in large errors. This explains the
relatively large uncertainty of these first measurements. Later
measurements showed that after working continuously for about a
month the instability disappeared. We obtained the result Cu-2 when
the counter had stabilised after this prolonged period of continuous
work.

\begin{figure}[htb]
\begin{center}\vspace{-0.2cm}
\includegraphics[width=10cm]{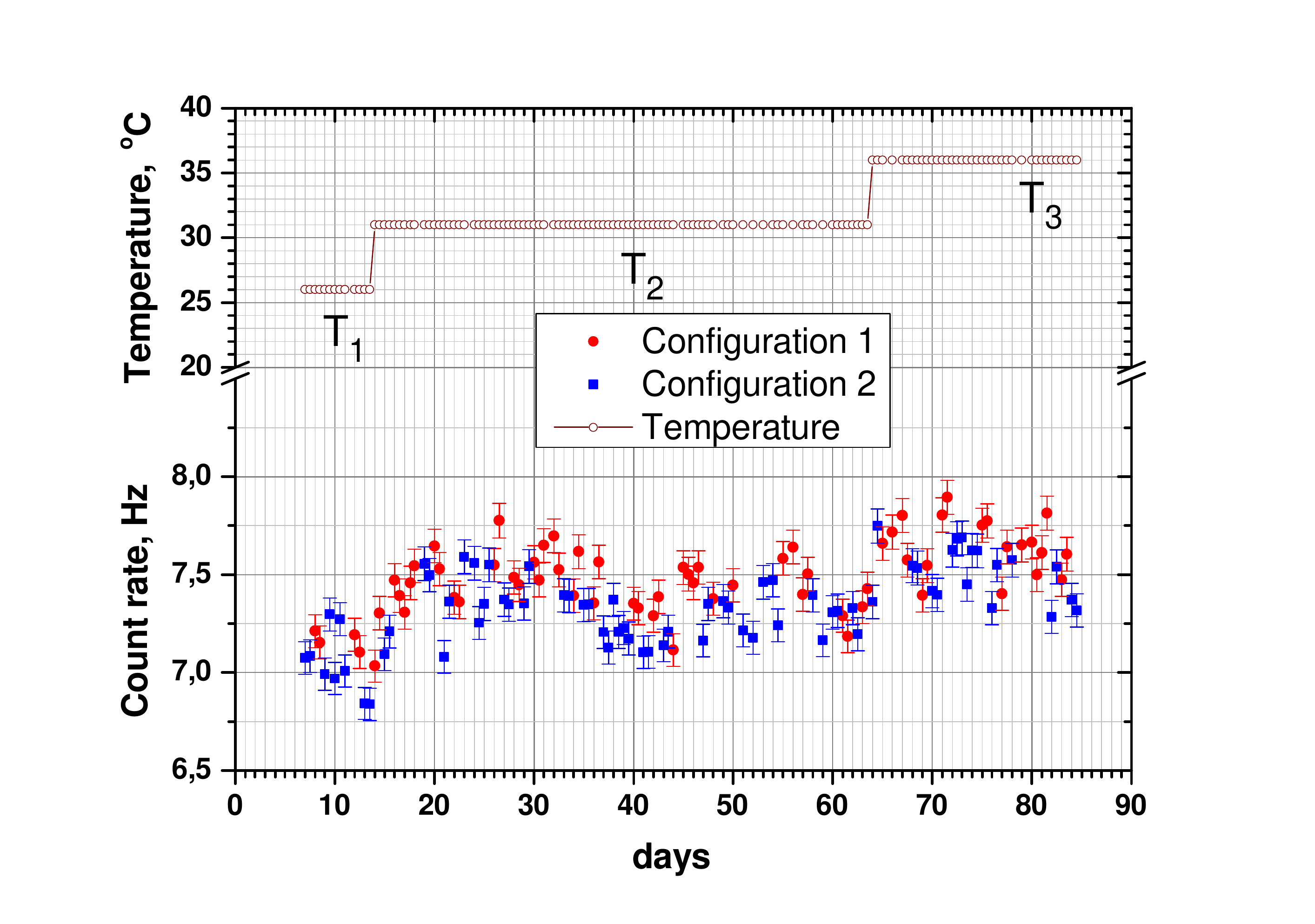}
\caption{\label{fig4} Results of measurements obtained using a
counter with a copper cathode at different temperatures.}
\end{center}
\end{figure}

\begin{figure}[htb]
\begin{center}\vspace{-0.2cm}
\includegraphics[width=10cm]{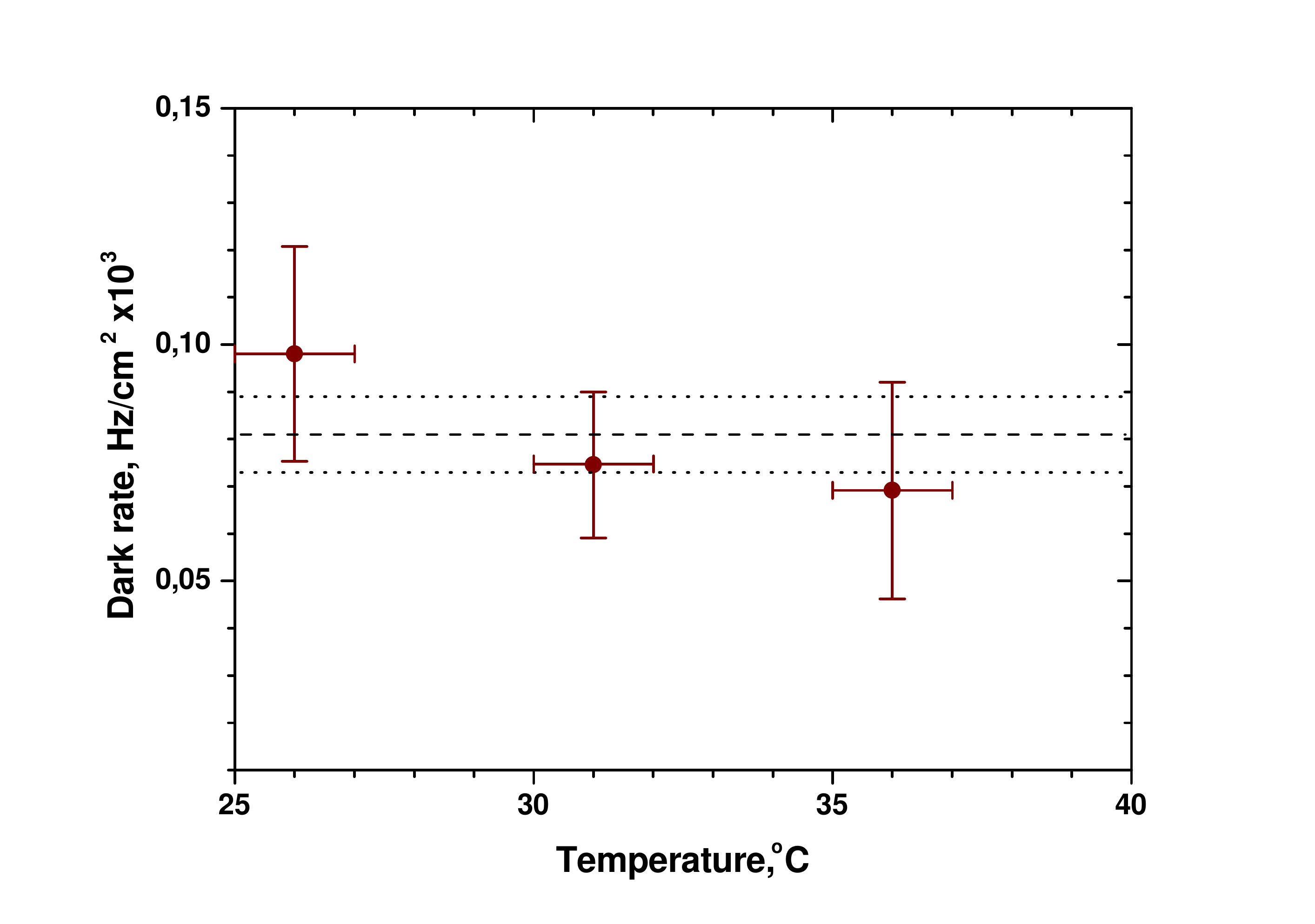}
\caption{\label{fig5} The distributions of events at $26^{o}$C,
$31^{o}$C and $36^{o}$C. The dashed line is the average value
obtained for all measurements, and the dotted lines represent $\pm$
1$\sigma$ levels from the average value.}
\end{center}
\end{figure}

After selecting for ``true'' pulses, we obtained for $r_{MCC} =
R_{MCC}/A_{cath}$:     $(0.98 \pm 0.22)\cdot 10^{-4} Hz/cm^2$,
$(0.75 \pm 0.15)\cdot 10^{-4} Hz/cm^2$ and $(0.69 \pm 0.23)\cdot
10^{-4} Hz/cm^2$ for the temperatures $26^{o}$C, $31^{o}$C and
$36^{o}$C, respectively (figure~\ref{fig5}). The fact that the count
rate does not exhibit a clear increase with temperature can be taken
as evidence that there is a negligible contribution from thermionic
emission. The average value obtained for all these temperatures is
$r_{MCC} = (0.81 \pm 0.08)\cdot 10^{-4} Hz/cm^2$.

\begin{figure}[htb]
\begin{center}\vspace{-0.2cm}
\includegraphics[width=10cm]{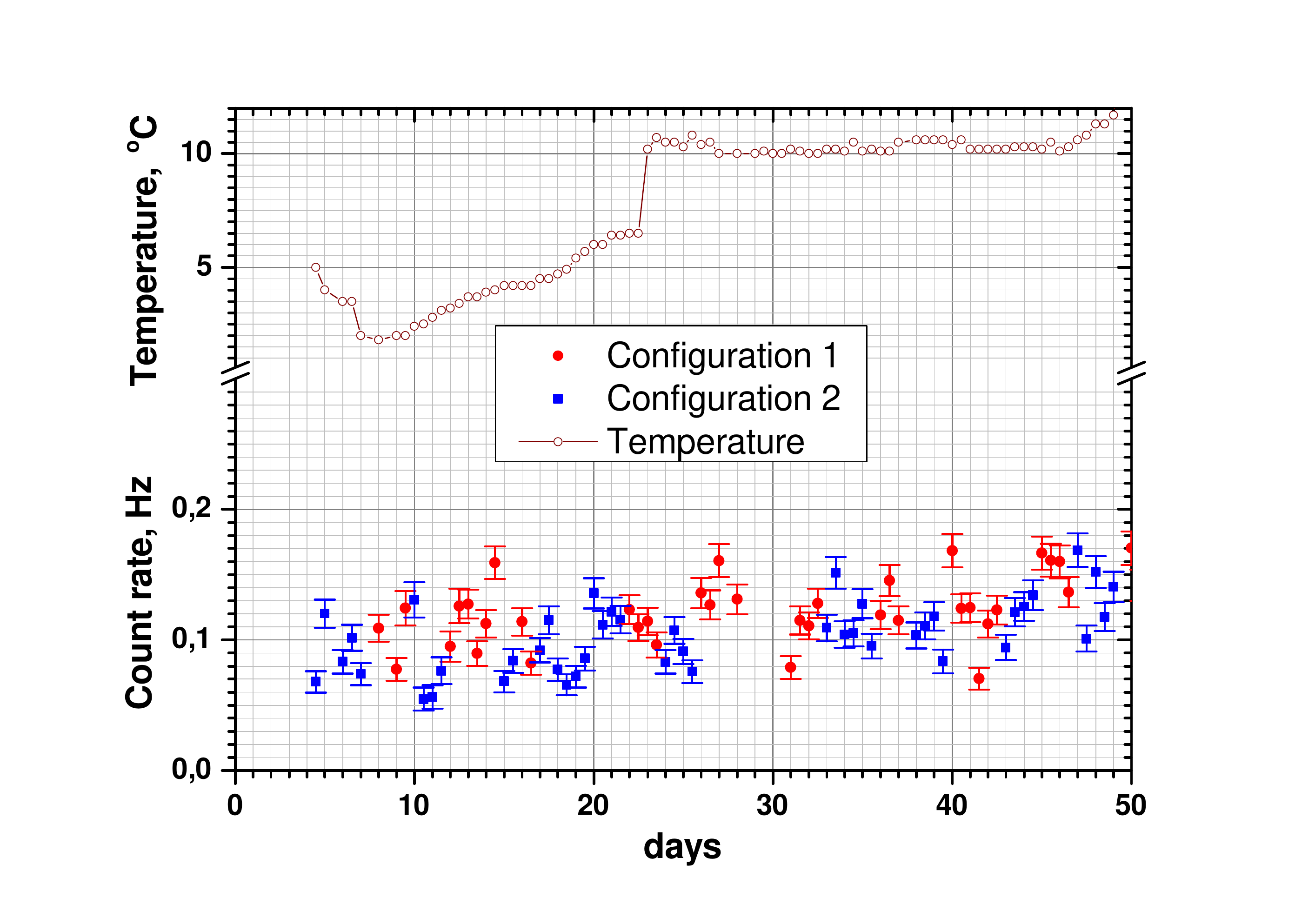}
\caption{\label{fig6} Results of measurements obtained using a
counter with an aluminium cathode.}
\end{center}
\end{figure}

We obtained the limit Al-1 from the measurements presented in
figure~\ref{fig6}, where we assumed that the main contribution to
the rate is due to the end effect.

The average value found for all the measurements presented in
figure~\ref{fig6}: $r_{MCC} = (0.8 \pm 0.25)\cdot 10^{-5} Hz/cm^2$.
This is the lowest result obtained to date.

\begin{figure}[htb]
\begin{center}\vspace{-0.2cm}
\includegraphics[width=10cm]{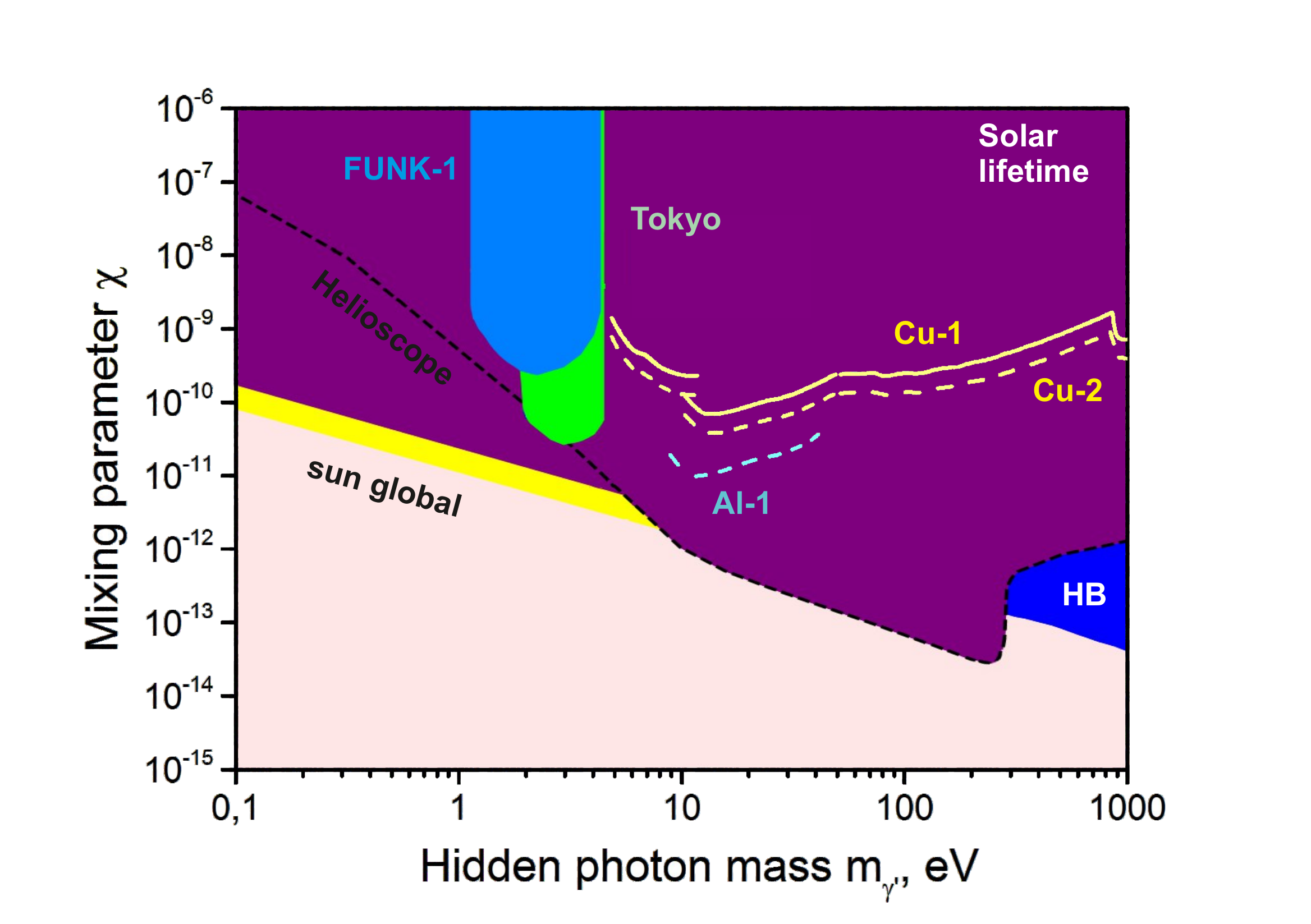}
\caption{\label{fig7} Limits at a 95\% CL obtained from the series
of measurements Cu-1, Cu-2 and Al-1. Here the Tokyo limits are
from~\cite{2} and the FUNK-1 limits from~\cite{10}.}
\end{center}
\end{figure}

The limits obtained at a 95\% confidence level (CL) are presented in
figure~\ref{fig7}. Using \eqref{eq:7} to account for the systematic
uncertainties would substantially decrease the limits presented in
this figure. The question of the relevant systematic uncertainty is
now under further study. If we were to assume that electrons emitted
from the cathode wires also contribute to the count rate,
\textit{i.e.} if we were to take into account our systematic errors,
the curves for Cu-1, Cu-2 and Al-1 in figure~\ref{fig7} would be
lower. An advanced surface treatment could further improve the
sensitivity of the measurements.

We have made good progress using a counter with an aluminium
cathode. Further improvements can be made by using Ni or Pt, which
have higher work functions, for the cathode of the counter. Figure 7
shows that our limits are higher than the limits given by the solar
lifetime, so we still need to make significant improvements to go
beyond these limits. However one should also take into account
possible uncertainties due to two relevant processes: the emission
of hidden photons inside the Sun and the emission of electrons from
the metal surfaces. There is also a significant difference between
volume and surface detectors. In the former, HPs are presumed to be
detected by their interactions with valence electrons, while in the
latter, detection is due to the interaction with free electrons in
the degenerate electron gas. Our detector is a surface detector, so
the limits obtained by this technique are related to the interaction
of HPs with the free electrons in the degenerate electron gas of the
metal.

\acknowledgments

The work was funded by the Federal Agency for Scientific
Organizations, Russia. The authors would like to thank Enago
(www.enago.com) for the English language review.

\end{document}